\documentclass[showpacs, prepintnumber,amssymb, nobibnotes, aps, prd]{revtex4}
\usepackage{graphicx}
\usepackage{dcolumn}
\usepackage{bm}

\begin{document}

\title{S-WAVE SCATTERING OF FERMION REVISITED 
TAKINFG INTO ACCOUNT THE EFFECT OF ANOMALY} 

\author{Anisur Rahaman}
\email{anisur.rahman@saha.ac.in}
\affiliation{Durgapur Govt. College, Durgapur - 713214, Burdwan, West Bengal, INDIA}

\date{\today}
\begin{abstract}A model where a Dirac fermion is coupled to 
 background dilaton
field is considered to study s-wave scattering of fermion by a
back ground dilaton black hole. It is found that the anomaly plays
a dual role on information loss scenario. Some time it protects
the fermion from the dangerous information loss problem and some
times it itself through it up towards that danger.
\end{abstract}
\maketitle
\section{introduction}
In recent years there has been a renewed interest in the physics
of information loss \cite{HAW, HOOFT, GARF, AHN, STRO, SUS, GIDD, GIDD1,
 CALL,  MIT, ARINF}. Matter falling into the black holes
carries some information with it. That becomes inaccessible to the
rest of the world and a problem arises when the black hole
evaporates through Hawking radiation. It is a controversial issue
whether or not quantum coherence would be maintained during the
formation and subsequent evaporation of a black hole.
Hawking initially suggested that the process could not preserve
information and unitarity failed to be maintained \cite{HAW}. It
was an indication of a new level of unpredictability in the realm
of quantum mechanics induced by gravity. There were spectrum of
contemporary opinions that went against Hawking's suggestion and
the issue gradually scattered in favor of preservation of
information \cite{HOOFT, GARF}.
Few years ago, Hawking himself moved away from his previous
suggestion and opined that quantum gravity interaction did not
lead to any loss of information. So there lies no problem to
maintain quantum coherence during the formation and subsequent
evaporation of the black hole \cite{HAW1}.
In spite of that, Hawking radiation effect on fermion information
loss problem is not well understood. Even now it has
been standing as a controversial issue \cite{AHN}. It is fair to admit that
information loss scenario in connection with fermion is not well
settled. In fact the nature of the problem is so mysterious that
there is the possibility of occurrence of information preservation
as well as non preservation which we would like to explore through
the s wave scattering of fermion off dilaton black hole.

Though it is very difficult to give a general description of this
problem, there have been attempts in studying such problem in its
full complexity through the  s-matrix description of such event
\cite{HOOFT}. Some less complicated models are around us and these
models are found to be very useful to study this paradox
\cite{STRO, SUS, GIDD, GIDD1, CALL}. Even in the presence of
gravitational anomaly a systematic description of scattering of
chiral fermion off dilaton gravity are found to be possible
through this type of model \cite{MIT, ARINF}. Another advantage
one can have from these studies is that there is a room to take
the effect of anomaly into consideration and that of course shows
a way to study the effect of anomaly on information loss issue. It
is worth mentioning at this stage that these simplified models did
not capture the detailed physics of black hole however those
models contain the information loss paradox in a significant way
\cite {STRO, SUS, MIT}.  Only the s-wave scattering of fermion
incident on the extremal charged black hole is considered in these
models and it is known that angular momentum coordinate becomes
irrelevant in this situation and a two dimensional effective
action is found to be resulted out \cite{GIDD, GIDD1}. From the
previous studies \cite{STRO, SUS, GIDD, GIDD1, CALL}, it was found
that information loss did not stood as a problem for Dirac
fermion. However we would like to argue that the issue is not
always so straightforward. In this context we would
like to mention the work available in \cite{MIT} and the recent
work of author \cite{ARINF} where the same issue was considered
for chiral fermion. In \cite{MIT}, we find a disastrous role of
anomaly where as in \cite{ARINF} we find it's novel role to
protect the model from that disaster. We should mention at this
stage that the conclusion drawn in \cite{MIT}, is correct indeed.
The only possible candidate that leads to the alteration of the
information scenario of chiral fermion is the anomaly structure of
the model because the two models differ only in their anomaly
structures. The present work is to re-investigate the effect of
anomaly in connection with the information loss problem through
the scattering of Dirac fermion off dilation black hole and to
strengthen the issue that anomaly really does have the potential
to alter the information loss scenario.

\section{Two dimensional effective model for studying s-wave scattering}
To this end we consider a model where Dirac fermion is coupled to
a background dilaton field $\Phi$. Of course, electromagnetic
background is taken into consideration. For sufficiently low
energy incoming fermion, the scattering of s-wave fermion incident
on a charge dilaton black hole can be described by the action
\begin{equation}
{\cal S}_f = \int d^2x[i\bar\psi\gamma^\mu[\partial_\mu +
ieA_\mu]\psi - {1\over 4} e^{-2\Phi(x)}F_{\mu\nu}F^{\mu\nu}].
\label{EQ1}
\end{equation}
Here e has one mass dimension. The indices $\mu$ and $\nu$ takes
the values $0$ and $1$ in $(1+1)$ dimensional space time. The
dilaton field $\Phi$ stands as a non dynamical back ground. It
completes its role in this model making the coupling constant a
position dependent one. Let us now define $g^2(x) = e^{2\Phi(x)}$.
We will choose a particular dilaton background motivated by the
linear dilaton vacuum of $(1+1)$ dimensional gravity like the
other standard cases \cite{GIDD, GIDD1, CALL, STRO, SUS, MIT,
ARINF}. Therefore, $\Phi(x) = -x^1$, where $x^1$ is space like
coordinate. The region $x^1 \to\ + \infty$, corresponds to exterior
space where the coupling $g^2(x)$ vanishes and the fermion will be
able to propagate freely. However, the region where $x^1 \to
-\infty$, the coupling constant will diverge and it is analogous
to infinite throat in the interior of certain magnetically charged
black hole.

The equation (\ref{EQ1}) is obtained from the action
\begin{equation}
S_{AF} = \int d^2\sigma\sqrt{g}[R + 4(\nabla\phi)^2 + {1\over
{Q^2}} - {1\over 2}F^2 + i \bar\psi D\!\!\!/\psi], \label{EQ2}
\end{equation}
for sufficiently low energy incoming fermion and negligible
gravitational effect \cite{STRO}. Here $D_\mu=\partial_\mu +
eA_\mu$. It is a two dimensional effective field theory of dilaton
gravity coupled to fermion. Here $\Phi$ represents the scalar
dilaton field and $\psi$ is the charged fermion. Equation
(\ref{EQ2}) was derived viewing the throat region of a four
dimensional dilaton black hole as a compactification from four to
two dimension \cite{GARF, GIDD, STRO}. Note that, in the extremal
limit, the geometry is completely non-singular and there is no
horizon but when a low energy particle is thrown into the
non-singular extremal black hole, it produces a singularity and an
event horizon. In this context, we should mention that the
geometry of the four dimensional dilaton black hole consists of
three regions \cite{GARF, STRO, SUS, GIDD, GIDD1}. First one is
the asymptotically flat region far from the black hole. As long as
one proceed nearer to the black hole the curvature begins to rise
and finally enters into the mouth region (the entry region to the
throat). Well into the throat region, the metric is approximated
by the flat two dimensional Minkowsky space times the round metric
on the two sphere with radius Q and equation (\ref{EQ2}) results.
The dilaton field $\Phi$ indeed increases linearly with the proper
distance into the throat.

We will start our analysis with the bosonization of the theory.
The advantage of the bosonized version is that a one loop
correction automatically enter within the model. In order to
bosonize the theory  we need to integrate out both the the left
handed as well as the right handed part of the fermion one by one
and express the fermionic determinant in terms of scalar field and
anomaly enters into the theory. So the tree level bosonized theory
gets the effect of anomaly during the process.  Of course,
bosonization can be done keeping  the gauge symmetry intact.
However we are interested to study the effect of anomaly on the
information loss scenario of Dirac fermion. So anomaly has been
taken into consideration.
With the anomaly used in the study of non anomalous Schwinger
model (some times it is termed as non confining Scgwinger model)
\cite{AR, AR1} the bosonized action reads
\begin{equation}
{\cal L}_B = {1\over 2}\partial_\mu\phi \partial^\mu\phi -
e\tilde\partial_\mu\phi A^\mu + {1\over 2}ae^2A_{\mu}A^{\mu} -
{1\over 4}e ^{2\Phi(x)}F_{\mu\nu}F^{\mu\nu}. \label{LBH}
\end{equation}
Here $\phi$ represents a scalar field and $\tilde\partial_\mu$ is
the dual to $\partial_\mu$. $\tilde\partial_\mu$ is defined by
$\tilde\partial_\mu=\epsilon_{\mu\nu}\partial^\nu$. The lagrangian
(\ref{LBH}) reminds us the anomalous (non-confining) vector
Schwinger model because for $\Phi(x)=0$, it maps on to the
non-confining Schwinger model \cite{AR, AR1}.

 The $U(1)$ current in this
situation is
\begin{equation}
J_\mu = -e\epsilon_{\mu\nu}\partial^\nu\phi + ae^2A_\mu
\end{equation}
and it is not conserved, i.e., $\partial_\mu J^\mu \neq 0$ which
was of preserving character within the descriptions available in
\cite{GIDD, GIDD1, STRO, SUS} and the current in those situations
was $J_\mu= -e\epsilon_{\mu\nu}\partial^\nu\phi$. The new setting
considered here to show role of anomaly on the information loss
scenario as has been already mentioned.
\section{Hamiltonian analysis of the model}
\subsection{Bosonized lagrangian amd constrained hamiltonian}
It is now necessary to carry out the Hamiltonian analysis of the
theory to observe the effect of the presence of dilaton field on
the equations of motion. From the standard definition the
canonical momenta corresponding to the scalar field $\phi$, and
the gauge fields $A_0$ and $A_1$ are found out:
\begin{equation}
\pi_\phi = \phi' - aA_1\label{MO1}
\end{equation}
\begin{equation}
\pi_0 = 0,\label{MO2}
\end{equation}
\begin{equation}
\pi_1 = e^{-2\phi(x)}(\dot A_1 - A_0')={1\over {g^2}}(\dot A_1 -
A_0).\label{MO3}
\end{equation}
Here $\pi_\phi$, $\pi_0$ and $\pi_1$ are the momenta corresponding
to the field $\phi$, $A_0$ and $A_1$. Using the above equations,
it is straightforward to obtain the canonical hamiltonian through
a Legendre transformation. The canonical hamiltonian is found out
to be
\begin{eqnarray}
{\cal H} &=& {1\over 2}(\pi_\phi +eA_1)^2 + {1\over
2}e^{2\phi(x)}\pi_1^2 + {1\over 2}\phi'^2 + \pi_1A_0' -eA_0\phi' \nonumber\\ &-&
{1\over 2}ae^2(A_0^2 - A_1^2).\label{CHAM}\end{eqnarray} 
Though we
find an explicit space dependence in the hamiltonian (\ref{CHAM})
through the dilaton field $\Phi(x)$, it has no time dependence. So
it is preserved in time. 
\subsection{Constrained analysis and theoretical spectra}
Equation (\ref{MO2}) is the familiar
primary constraints of the theory. Therefore, it is necessary to
write down an  effective hamiltonian:
\begin{equation}
{\cal H}_{eff} = {\cal H}_C + u\pi_0
\end{equation}
where $u$ is an arbitrary Lagrange multipliers. The primary
constraints (\ref{MO1}) has to be preserve in order to have a
consistent theory. The preservation of the constraint (\ref{MO2}),
leads to the Gauss law of the theory as a secondary constraint:
\begin{equation}
G = \pi_1' + 2e\phi' +  ae^2A_0 \approx 0. \label{GAUS}
\end{equation}
The preservation of the constraint (\ref{GAUS}) though does not
give rise to any new constraint it fixes the velocity $u$ which
comes out to be
\begin{equation}
u =A'_1. \label{VEL}
\end{equation}
We, therefore, find that the phase space of the theory contains
the following two second class constraints.
\begin{equation}
\omega_1 = \pi_0 \approx 0, \label{CON1}
\end{equation}
\begin{equation}
\omega_2 = \pi_1' + 2e\phi' +  ae^2A_0 \approx 0,\label{CON2}
\end{equation}
Both the constraints (\ref{CON1}) and (\ref{CON2}) are weak
conditions up to this stage. When we impose these constraints
strongly into the canonical hamiltonian (\ref{CHAM}), the
canonical hamiltonian gets simplified into the following.
\begin{eqnarray}
{\cal H}_{red} &=& {1\over 2}(\pi_\phi + eA_1)^2 + {1\over
{2ae^2}}(\pi'_1 + e\phi')^2 + {1\over 2}e^{2\Phi(x)}\pi_1^2 
\nonumber\\ 
&+&{1\over 2}\phi'^2 + {1\over 2}ae^2A_1^2.
\label{RHAM}\end{eqnarray} 
$H_{red}$ given in equation
(\ref{RHAM}), is generally known as reduced Hamiltonian. According
to Dirac \cite{DIR}, Poisson bracket gets invalidate for this reduced
Hamiltonian . This reduced Hamiltonian however remains
consistent with the Dirac brackets which is defined by
\begin{eqnarray}
& &[A(x), B(y )]^* = [A(x), B(y)] \nonumber \\ 
&-&\int[A(x), \omega_i(\eta)]
C^{-1}_{ij}(\eta, z)[\omega_j(z), B(y)]d\eta dz, \label{DEFD}
\end{eqnarray}
where $C^{-1}_{ij}(x,y)$ is given by
\begin{equation}
\int C^{-1}_{ij}(x,z) [\omega_j(z), \omega_k(y)]dz =\delta(x-y)
\delta_{ik}. \label{INV}
\end{equation}
For the theory under consideration \noindent $C_{ij}(x,y) =$
\begin{equation}
ae^2 \pmatrix {0 & -\delta(x-y) \cr \delta(x-y) & o \cr}
\label{MAT}
\end{equation}

Here $i$ and $j$ runs from $1$ to $2$ and $\omega$'s represent the
constraints of the theory. With the definition (\ref{DEFD}), we
can compute the Dirac brackets between the fields describing the
reduced Hamiltonian $H_{red}$. The Dirac brackets between the
fields $A_1$, $\pi_1$, $\phi$ and $\pi_\phi$ are required to
obtain the theoretical spectra (equations of motion):
\begin{equation}
[A_1(x), A_1(y)]^* = 0 = [\pi_1(x), \pi_1(y)]^* \label{DR1}
\end{equation}
\begin{equation}
[A_1(x), \pi_1(y)]^* = \delta(x-y),\label{DR2}
\end{equation}
\begin{equation}
[\phi(x), \Phi(y)]^* = 0 =[\pi_\phi(x), \pi_\phi(y)]^* \label{DR3}
\end{equation}
\begin{equation}
[\phi(x), \Pi_\phi(y)]^* = \delta(x-y) \label{DR4}
\end{equation}
We are now in a position to find out the equations of motion from
the reduced hamiltonian(\ref{RHAM}). With the use of Dirac
Brackets (\ref{DR1}), (\ref{DR2}), (\ref{DR3}) and (\ref{DR4},
Heisenberg's equation of motion for the reduced
hamiltonian(\ref{RHAM}) leads to the following four first order
equations.
\begin{equation}
\dot A_1= e^{2\Phi}\pi_1 -{1\over {ae^2}}(\pi_1'' + e\phi'')
,\end{equation}

\begin{equation}
\dot\phi = \pi_\phi + eA_1 ,
\end{equation}

\begin{equation}
\pi_\phi = {{a+1}\over a}\phi'' + {1\over {ae}}\pi_1''
,\end{equation}

\begin{equation}
\dot\pi_1 = -e\pi_\phi - (a+1)e^2A_1. \end{equation}
A little
algebra converts the four first order equations into the following
two second order Klein-Gordon equations:

\begin{equation}
[\Box + (1+a)e^2e^{2\Phi(x)}]\pi_1 = 0, \label{SP1}
\end{equation}
\begin{equation}
\Box[\pi_1 + e(1+a)\phi] = 0. \label{SP2}
\end{equation}

The equation (\ref{SP1}), represents a massive boson with square
of the mass $m^2 =(1+a)e^2e^{2\Phi(x)}$. Here $a$ must be greater
than $-1$ in order to have the mass of the boson a physical one.
Equation (\ref{SP2}) however describes a massless boson. The presence
of this massless boson has a disastrous role on the information scenario
which will be uncovered in the following section. 

\section{Discussion}
Let us concentrate into the theoretical spectra. Mass of this boson as 
appeared in (\ref{SP1}) is not constant in this particular situation. It contains a position dependent factor $g^2=e^{2\Phi(x)}$ where $\Phi = -x^1$, for the background motivated by the linear
dilaton vacuum of $(1+1)$ dimensional gravity. Therefore, $m^2 \to
\ + \infty$ when $x^1 \to\ - \infty$ and $m^2 \to \ 0$ when $x^1
\to\ + \infty$. Thus mass of the boson is found to be increased
indefinitely in the negative $x^1$ direction which implies that
any finite energy contribution must be totally reflected and an
observer at $x^1 \to \infty$ will recover all information. To be
more precise, mass will vanish near the mouth (the entry region to
the throat) but increases indefinitely as one goes into the throat
because of the variation of this space dependent factor $g^2$.
Since massless scalar is equivalent to massless fermion in $(1+1)$
dimension, we can conclude that a massless fermion proceeding into
the black hole will not be able to travel an arbitrarily long
distance and will be reflected back with a unit probability. So
there is no threat regarding information loss from the massive
sector of the theory. However an uncomfortable situation appears
when we look carefully towards the massless sector as described by
the equation (\ref{SP2}).
This remains massless irrespective of its position.  So, this
fermion will be able to travel within the black hole with out any
hindrance and an observer at $x^1 \to \infty$ will never find this
fermion with a back ward journey. Thus the problem of information
loss becomes very real with this setting. Note that in the similar
type of studies \cite{STRO, SUS, GIDD, GIDD1} where the setting
was such that it was anomaly free the problem of information loss
did not occur. The result of the present work though leads to an
uncomfortable situation, it is of particular importance and
significance in connection with  the role of anomaly on the
problem of information loss and a careful look revels that it has
appeared just as the out come of the allowance of the anomaly
within the model. In the previous work of the author \cite{ARINF},
one finds the reverse role of anomaly for a particular setting
where anomaly stood as a protector of the chiral fermion from the
danger of information loss. The result obtained from the works
\cite{ STRO, SUS, GIDD, GIDD1} for Dirac fermion and from the work
\cite{ARINF} for chiral fermion looks comfortable since
information loss problem did not appear with the settings
considered therein and it is consistent with the Hawking's recent
suggestion and as well as with the standard belief. However the
results of the work \cite{MIT} for chiral fermion and the result
of the present work for Dirac fermion do not look so, since
information loss problem can not be avoided with the settings
considered within these and these two go against Hawking's
recent suggestion and of course against the standard belief. It
has to be remember that we are interested in the issue related to the
role of anomaly rather than the result we are getting and a 
comparative studies of
the previous works \cite {STRO, SUS, GIDD, GIDD1, MIT, ARINF} and
the present work confirm that the anomaly is playing a dual role
on the problem of information loss related to Dirac fermion as well 
as chiral
fermion. Some time it emerges out as a protector from the danger of
information loss and some time it itself throw it up towards the
same. It does not come as a great surprise. The crucial role of
anomaly was noticed earlier in the description of quantum
electrodynamics and quantum chiral electrodynamics \cite{AR, AR1,
BEL, JR, KH, PM, MG, FLO}. A famous instance in this context is
the removal of the long suffering of the chiral electrodynamics
from the non unitarity problem \cite{JR}
To conclude this we would like to mention that anomaly sometimes
brings in a disaster and sometimes it itself stands as a saver of
the same. However one question may be asked which result is
acceptable? It is fair to admit that there is no specific answer.
However some comment can be made. If preservation of information
is considered to be the acceptable then the anomaly that
correspond to the preservation of information has to be chosen.
The situation would be reverse indeed if the non-preservation of
information is considered to be acceptable. So to study the
information loss through this type of model it is important to fix
up the setting in order to get a desired result.\\

\noindent{\bf Acknowledgment}: It is a pleasure to thank the
Director, Saha Institute of Nuclear Physics and the Head of the
Theory Group of Saha Institute of Nuclear Physics, Kolkata for
providing working facilities.


\begin{thebibliography}{the}
\bibitem {HAW} S. W. Hawking, Commun. Math. Phys, {\bf 43} (1975) 199
\bibitem {HOOFT} G. 't Hooft, Nucl. Phys. {\bf B335} (1990) 138
\bibitem {GARF} D. Garfinkle, G. Horowitz and A. Strominger, Phys. Rev. {\bf D43} (1991) 3140
\bibitem {HAW1} S. W. Hawking, Phys. Rev {\bf D72} (2005) 084013
\bibitem {AHN} D. Ahn, quant-Ph/0604080
\bibitem {STRO} M. Alford and A. Strominger, Phys. Rev. Lett. {\bf 69} (1992) 563
\bibitem {SUS} A. Peet, L. Susskind and L. Thorlacius, Phys. Rev. {\bf D46} (1992) 3435
\bibitem {GIDD} S. Giddings and A. Strominger, Phys. Rev. {\bf D46} (1992) 627
\bibitem {GIDD1} T. Banks, A. Dabholkar, M. Douglas and M. O'Loughlin, Phys.
Rev. {\bf D45} (1992) 3607
\bibitem {CALL} C. Callan, S. Giddings, J. Hervey and A. Strominger, Phys. Rev.
{\bf D45} (1992) R1005
\bibitem {MIT} A. Ghosh and P. Mitra, Phys. Rev. {\bf D50} (1994) 7389
\bibitem {ARINF} A. Rahaman, ArXiv: 0901.3693, to appear in Mod. Phys. Lett. A
\bibitem {DIR} P. A. M. Dirac, Lectures on Quantum Mechanics(Yeshiva Univ. Press New York, 1964)
\bibitem {AR} P. Mitra and A. Rahaman, Ann. Phys. (N.Y.) {\bf249} (1996) 34
\bibitem {AR1} A. Rahaman, Int. Jour. Mod. Phys. {\bf A21 } (2006) 1251
\bibitem {BEL} S. Bellucci, M. F. L. Golterman and D. N. Petcher, Nucl. Phys.
{\bf B326} (1989) 307
\bibitem {JR} R Jackiw and R Rajaraman, Phys. Rev. Lett. {\bf 54} (1985) 1219
\bibitem {KH} K. Harada, Phys. Rev. Lett. {\bf 64} (1990) 139
\bibitem {PM} P. Mitra, Phys. Lett {\bf B284} (1992) 23
\bibitem {MG} S. Ghosh and P. Mitra, Phys. Rev. {\bf D44} (1991) 1332
\bibitem {FLO} R. Floreanini, R. Jackiw, Phys. Rev. Lett. {\bf 59}(1987) 1873

\end{thebibliography}
\end{document}